%% file: main.tex
\documentclass[twocolumn]{aastex63}
\usepackage{amsmath}
\usepackage{amssymb}
\usepackage{amsthm}
\usepackage{booktabs}
\usepackage{graphicx}
\usepackage[colorinlistoftodos]{todonotes}
\usepackage{natbib}

\usepackage{lineno}
\usepackage{subfigure}
\usepackage{soul}
\usepackage{verbatim}

\setwatermarkfontsize{0.75in}
\definecolor{darkgreen}{rgb}{0.05,0.3,0.05}

\newcommand{\ZipperNet}{ZipperNet}

\begin{document}

\vspace*{-\headsep}\vspace*{\headheight}
{\footnotesize \hfill FERMILAB-PUB-21-392-E-SCD}\\
\vspace*{-\headsep}\vspace*{\headheight}
{\footnotesize \hfill DES-2021-0688}

\title{DeepZipper: A Novel Deep Learning Architecture for Lensed Supernovae Identification}
%\title{DeepLensNova I: Searching for Lensed Supernovae with \ZipperNet}
% DeepSNe
% DeepTransients
% DeepLenSNe
% DeepChange
% DeepCadence
% DeepLenstronomSNe
% DeepLensNovae

\input{authors}

\begin{abstract}
\input{abstract}
\end{abstract}


%\linenumbers

\section{Introduction}
\label{sec:introduction}
\input{introduction}

\section{Methods}
\label{sec:methods}
\input{methods}

\section{Results}
\label{sec:results}

\input{results}

\section{Discussion}
\label{sec:discussion}
\input{discussion}

\section{Conclusion}
\label{sec:conclusion}

\input{conclusion}

\section*{Acknowledgments}
\input{acknowledgements}

\software{
\texttt{astropy} \citep[]{astropy},
\texttt{deeplenstronomy} \citep[]{deeplenstronomy}, 
%\texttt{h5py} \citep[]{h5py},
\texttt{lenstronomy} \citep[]{lenstronomy},
\texttt{matplotlib} \citep[]{matplotlib},
\texttt{numpy} \citep[]{numpy}, 
\texttt{pandas} \citep[]{pandas},
\texttt{PlotNeuralNet} \citep[]{network_vis}, 
\texttt{PyTorch} \citep[]{pytorch},
\texttt{Scikit-Learn} \citep[]{sklearn},
\texttt{scipy} \citep[]{scipy}.
}

\clearpage

\bibliographystyle{yahapj_twoauthor_arxiv_amp}
\bibliography{main}
\end{document}

%% file: authors.tex
\correspondingauthor{Robert Morgan}
\email{robert.morgan@wisc.edu}

\author[0000-0002-7016-5471]{R.~Morgan}
\affil{Physics Department, University of Wisconsin-Madison, 1150 University Avenue Madison, WI  53706, USA}
\affil{Fermi National Accelerator Laboratory, P. O. Box 500, Batavia, IL 60510, USA}
\affil{Legacy Survey of Space and Time Corporation Data Science Fellowship Program, USA}

\author[0000-0001-6706-8972]{B.~Nord}
\affil{Fermi National Accelerator Laboratory, P. O. Box 500, Batavia, IL 60510, USA}
\affil{Department of Astronomy and Astrophysics, University of Chicago, Chicago, IL 60637, USA}
\affil{Kavli Institute for Cosmological Physics, University of Chicago, Chicago, IL 60637, USA}

\author[0000-0001-8156-0429]{K.~Bechtol}
\affil{Physics Department, University of Wisconsin-Madison, 1150 University Avenue Madison, WI  53706, USA}
\affil{Legacy Survey of Space and Time, 933 North Cherry Avenue, Tucson, AZ 85721, USA}

\author[0000-0001-7282-3864]{S.~J.~Gonz\'{a}lez}
\affil{Physics Department, University of Wisconsin-Madison, 1150 University Avenue Madison, WI  53706, USA}

\author[0000-0002-3304-0733]{E.~Buckley-Geer}
\affil{Fermi National Accelerator Laboratory, P. O. Box 500, Batavia, IL 60510, USA}
\affil{Department of Astronomy and Astrophysics, University of Chicago, Chicago, IL 60637, USA}

\author[0000-0001-8211-8608]{A.~M\"oller}
\affil{Centre for Astrophysics \& Supercomputing, Swinburne University of Technology, Victoria 3122, Australia}

\author[0000-0002-0692-1092]{J.~W.~Park}
\affil{Kavli Institute for Particle Astrophysics and Cosmology, Department of Physics, Stanford University, Stanford, CA 94305, USA}
\affil{SLAC National Accelerator Laboratory, Menlo Park, CA 94025, USA}

\author{A.~G.~Kim}
\affil{Lawrence Berkeley National Laboratory, 1 Cyclotron Road, Berkeley, CA 94720, USA}

\author[0000-0003-3195-5507]{S.~Birrer}
\affil{Kavli Institute for Particle Astrophysics and Cosmology, Department of Physics, Stanford University, Stanford, CA 94305, USA}
\affil{SLAC National Accelerator Laboratory, Menlo Park, CA 94025, USA}

\author{M.~Aguena}
\affil{Laborat\'orio Interinstitucional de e-Astronomia - LIneA, Rua Gal. Jos\'e Cristino 77, Rio de Janeiro, RJ - 20921-400, Brazil}

\author[0000-0002-0609-3987]{J.~Annis}
\affil{Fermi National Accelerator Laboratory, P. O. Box 500, Batavia, IL 60510, USA}

\author[0000-0002-4900-805X]{S.~Bocquet}
\affil{Faculty of Physics, Ludwig-Maximilians-Universit\"at, Scheinerstr. 1, 81679 Munich, Germany}

\author[0000-0002-8458-5047]{D.~Brooks}
\affil{Department of Physics \& Astronomy, University College London, Gower Street, London, WC1E 6BT, UK}

\author[0000-0003-3044-5150]{A.~Carnero~Rosell}
\affil{Laborat\'orio Interinstitucional de e-Astronomia - LIneA, Rua Gal. Jos\'e Cristino 77, Rio de Janeiro, RJ - 20921-400, Brazil}

\author[0000-0002-4802-3194]{M.~Carrasco~Kind}
\affil{Center for Astrophysical Surveys, National Center for Supercomputing Applications, 1205 West Clark St., Urbana, IL 61801, USA}
\affil{Department of Astronomy, University of Illinois at Urbana-Champaign, 1002 W. Green Street, Urbana, IL 61801, USA}

\author[0000-0002-3130-0204]{J.~Carretero}
\affil{Institut de F\'{\i}sica d'Altes Energies (IFAE), The Barcelona Institute of Science and Technology, Campus UAB, 08193 Bellaterra (Barcelona) Spain}

\author{R.~Cawthon}
\affil{Physics Department, William Jewell College, Liberty, MO 64068 USA}

\author{L.~N.~da Costa}
\affil{Laborat\'orio Interinstitucional de e-Astronomia - LIneA, Rua Gal. Jos\'e Cristino 77, Rio de Janeiro, RJ - 20921-400, Brazil}
\affil{Observat\'orio Nacional, Rua Gal. Jos\'e Cristino 77, Rio de Janeiro, RJ - 20921-400, Brazil}

\author[0000-0002-4213-8783]{T.~M.~Davis}
\affil{School of Mathematics and Physics, University of Queensland,  Brisbane, QLD 4072, Australia}

\author[0000-0001-8318-6813]{J.~De~Vicente}
\affil{Centro de Investigaciones Energ\'eticas, Medioambientales y Tecnol\'ogicas (CIEMAT), Madrid, Spain}

\author{P.~Doel}
\affil{Department of Physics \& Astronomy, University College London, Gower Street, London, WC1E 6BT, UK}

\author{I.~Ferrero}
\affil{Institute of Theoretical Astrophysics, University of Oslo. P.O. Box 1029 Blindern, NO-0315 Oslo, Norway}

\author{D.~Friedel}
\affil{Center for Astrophysical Surveys, National Center for Supercomputing Applications, 1205 West Clark St., Urbana, IL 61801, USA}

\author[0000-0003-4079-3263]{J.~Frieman}
\affil{Fermi National Accelerator Laboratory, P. O. Box 500, Batavia, IL 60510, USA}
\affil{Kavli Institute for Cosmological Physics, University of Chicago, Chicago, IL 60637, USA}

\author[0000-0002-9370-8360]{J.~Garc\'ia-Bellido}
\affil{Instituto de Fisica Teorica UAM/CSIC, Universidad Autonoma de Madrid, 28049 Madrid, Spain}

\author{M.~Gatti}
\affil{Department of Physics and Astronomy, University of Pennsylvania, Philadelphia, PA 19104, USA}

\author[0000-0001-9632-0815]{E.~Gaztanaga}
\affil{Institut d'Estudis Espacials de Catalunya (IEEC), 08034 Barcelona, Spain}
\affil{Institute of Space Sciences (ICE, CSIC),  Campus UAB, Carrer de Can Magrans, s/n,  08193 Barcelona, Spain}

\author[0000-0002-3730-1750]{G.~Giannini}
\affil{Institut de F\'{\i}sica d'Altes Energies (IFAE), The Barcelona Institute of Science and Technology, Campus UAB, 08193 Bellaterra (Barcelona) Spain}

\author[0000-0003-3270-7644]{D.~Gruen}
\affil{Faculty of Physics, Ludwig-Maximilians-Universit\"at, Scheinerstr. 1, 81679 Munich, Germany}

\author{R.~A.~Gruendl}
\affil{Center for Astrophysical Surveys, National Center for Supercomputing Applications, 1205 West Clark St., Urbana, IL 61801, USA}
\affil{Department of Astronomy, University of Illinois at Urbana-Champaign, 1002 W. Green Street, Urbana, IL 61801, USA}

\author[0000-0003-0825-0517]{G.~Gutierrez}
\affil{Fermi National Accelerator Laboratory, P. O. Box 500, Batavia, IL 60510, USA}

\author{D.~L.~Hollowood}
\affil{Santa Cruz Institute for Particle Physics, Santa Cruz, CA 95064, USA}

\author[0000-0002-6550-2023]{K.~Honscheid}
\affil{Center for Cosmology and Astro-Particle Physics, The Ohio State University, Columbus, OH 43210, USA}
\affil{Department of Physics, The Ohio State University, Columbus, OH 43210, USA}

\author[0000-0001-5160-4486]{D.~J.~James}
\affil{Center for Astrophysics $\vert$ Harvard \& Smithsonian, 60 Garden Street, Cambridge, MA 02138, USA}

\author[0000-0003-0120-0808]{K.~Kuehn}
\affil{Australian Astronomical Optics, Macquarie University, North Ryde, NSW 2113, Australia}
\affil{Lowell Observatory, 1400 Mars Hill Rd, Flagstaff, AZ 86001, USA}

\author[0000-0003-2511-0946]{N.~Kuropatkin}
\affil{Fermi National Accelerator Laboratory, P. O. Box 500, Batavia, IL 60510, USA}

\author[0000-0001-9856-9307]{M.~A.~G.~Maia}
\affil{Laborat\'orio Interinstitucional de e-Astronomia - LIneA, Rua Gal. Jos\'e Cristino 77, Rio de Janeiro, RJ - 20921-400, Brazil}
\affil{Observat\'orio Nacional, Rua Gal. Jos\'e Cristino 77, Rio de Janeiro, RJ - 20921-400, Brazil}

\author[0000-0002-6610-4836]{R.~Miquel}
\affil{Instituci\'o Catalana de Recerca i Estudis Avan\c{c}ats, E-08010 Barcelona, Spain}
\affil{Institut de F\'{\i}sica d'Altes Energies (IFAE), The Barcelona Institute of Science and Technology, Campus UAB, 08193 Bellaterra (Barcelona) Spain}

\author[0000-0002-6011-0530]{A.~Palmese}
\affil{Department of Astronomy, University of California, Berkeley,  501 Campbell Hall, Berkeley, CA 94720, USA}

\author{F.~Paz-Chinch\'{o}n}
\affil{Center for Astrophysical Surveys, National Center for Supercomputing Applications, 1205 West Clark St., Urbana, IL 61801, USA}
\affil{Institute of Astronomy, University of Cambridge, Madingley Road, Cambridge CB3 0HA, UK}

\author{M.~E.~S.~Pereira}
\affil{Department of Physics, University of Michigan, Ann Arbor, MI 48109, USA}
\affil{Hamburger Sternwarte, Universit\"{a}t Hamburg, Gojenbergsweg 112, 21029 Hamburg, Germany}

\author[0000-0001-9186-6042]{A.~Pieres}
\affil{Laborat\'orio Interinstitucional de e-Astronomia - LIneA, Rua Gal. Jos\'e Cristino 77, Rio de Janeiro, RJ - 20921-400, Brazil}
\affil{Observat\'orio Nacional, Rua Gal. Jos\'e Cristino 77, Rio de Janeiro, RJ - 20921-400, Brazil}

\author[0000-0002-2598-0514]{A.~A.~Plazas~Malag\'on}
\affil{Department of Astrophysical Sciences, Princeton University, Peyton Hall, Princeton, NJ 08544, USA}

\author{K.~Reil}
\affil{SLAC National Accelerator Laboratory, Menlo Park, CA 94025, USA}

\author[0000-0001-5326-3486]{A.~Roodman}
\affil{Kavli Institute for Particle Astrophysics \& Cosmology, P. O. Box 2450, Stanford University, Stanford, CA 94305, USA}
\affil{SLAC National Accelerator Laboratory, Menlo Park, CA 94025, USA}

\author[0000-0002-9646-8198]{E.~Sanchez}
\affil{Centro de Investigaciones Energ\'eticas, Medioambientales y Tecnol\'ogicas (CIEMAT), Madrid, Spain}

\author[0000-0002-3321-1432]{M.~Smith}
\affil{School of Physics and Astronomy, University of Southampton,  Southampton, SO17 1BJ, UK}

\author[0000-0002-7047-9358]{E.~Suchyta}
\affil{Computer Science and Mathematics Division, Oak Ridge National Laboratory, Oak Ridge, TN 37831}

\author{M.~E.~C.~Swanson}
\affil{School of Physics and Astronomy, University of Southampton, Southampton, SO17 1BJ, UK}

\author[0000-0003-1704-0781]{G.~Tarle}
\affil{Department of Physics, University of Michigan, Ann Arbor, MI 48109, USA}

\author[0000-0001-7836-2261]{C.~To}
\affil{Center for Cosmology and Astro-Particle Physics, The Ohio State University, Columbus, OH 43210, USA}

%% file: abstract.tex
Large-scale astronomical surveys have the potential to capture data on large numbers of strongly gravitationally lensed supernovae (LSNe).
To facilitate timely analysis and spectroscopic follow-up before the supernova fades, an LSN needs to be identified soon after it begins. 
To quickly identify LSNe in optical survey datasets, we designed ZipperNet, a multi-branch deep neural network that combines convolutional layers (traditionally used for images)  with long short-term memory (LSTM) layers (traditionally used for time series).
We tested ZipperNet on the task of classifying objects from four categories -- no lens, galaxy-galaxy lens, lensed type Ia supernova, lensed core-collapse supernova -- within high-fidelity simulations of three cosmic survey data sets -- the Dark Energy Survey (DES), Rubin Observatory's Legacy Survey of Space and Time (LSST), and a Dark Energy Spectroscopic Instrument (DESI) imaging survey.
Among our results, we find that for the LSST-like dataset, ZipperNet classifies LSNe with a receiver operating characteristic area under the curve of 0.97, predicts the spectroscopic type of the lensed supernovae with 79\% accuracy, and demonstrates similarly high performance for LSNe 1-2 epochs after first detection.
We anticipate that a model like ZipperNet, which simultaneously incorporates spatial and temporal information, can play a significant role in the rapid identification of lensed transient systems in cosmic survey experiments.
% Strongly gravitationally lensed supernovae (LSNe) are uniquely powerful probes of the Universe's expansion rate today ($H_0$).
% for time-delay cosmography measurements of $H_0$ during the Rubin Observatory main survey operations.

%% file: introduction.tex
\begin{table*}
    \centering
    \begin{tabular}{|c|ccc|ccc|c|} \toprule
       Dataset  & Gain & Read Noise & Pixel Size & Exp. Time &  Seeing & Cadence & Reference \\ 
       & [e$^-$/count] & [e$^-$] & [arcsec] & [seconds] & [FWHM arcsec] &  [days] & \\ \midrule
       \textbf{DES-wide} & 6.083 & 7.0 & 0.263 & 90 &  $0.91 \pm 0.12$ & $ 220 \pm 191 $ & \citet{desdr1} \\
       \textbf{LSST-wide} & 2.3 & 10.0 & 0.2 & 30 &  0.71 & $12 \pm 5$ & \citet{lsst_table} \\
       \textbf{DES-deep} & 6.083 & 7.0 & 0.263 & 200 &  $1.04 \pm 0.44$ & $ 6 \pm 1$ & \citet{dessn} \\
       \textbf{DESI-DOT} & 6.083 & 7.0 & 0.263 & 60 &  $0.90 \pm 0.12$ & 3 & \citet{desdr1} \\
      \bottomrule
    \end{tabular}
    \caption{
    Summary of the instruments and observational procedures emulated in the simulated data sets: camera properties, observing conditions, and survey cadence.
    We show only the $i$ band for each property, because it illustrates the key discerning features between the datasets.
    The cadence displays the mean and standard deviation of the intra-band time separation. 
    The full cadence information and data quality properties are available in the \texttt{deeplenstronomy} input files, which accompany this work \citep{zenodo}.}
    \label{tab:datasets}
\end{table*}

Strong gravitational lensing is an expansive probe of both astrophysics and cosmology.
In systems with strong lensing, light from a background object (the source) is deflected by the gravitational potential of an interposed foreground object (the lens), producing characteristic features such as arcs, Einstein rings, and/or multiple images \citep[]{slreview}.
A key subclass of lenses, strongly lensed transients, have time-variable brightness in the source galaxy.
Due to the cosmological distances involved in strong lensing, the most common transient objects to observe lensed are quasars, which vary in brightness on the time scales of several years \citep[]{quasar_variability_2, quasar_variability}, and supernovae, which can reach peak brightness within days and then dim over the course of weeks to months \citep[]{sn_decay}.

One of the principle features of lensed transients is the time delay between arrival of photons that take different paths around the lenses.
Photons from multiple images of source objects travel different distances to Earth and experience different magnitudes of the gravitational potential due to the geometry of the lensing system.
Given the constant speed of light, the differences in the path lengths and gravitational potentials traversed by the photons produce an offset in the arrival times for photons that are emitted at the same point in the source's lightcurve.
Therefore, sources with time-varying brightness in these systems exhibit an observable time delay between the individual images of the lensed source \citep{tdcosmography}.

Strongly lensed transients are particularly useful for a wide range of investigations.
For example, the magnification of these background sources and their environments can reveal new information on the eruption and prevalence of these objects at earlier times in the Universe \citep{tdcmreview}.
Time-delay cosmography (TDC) is a technique that uses the time-varying brightness of compact systems that have undergone strong gravitational lensing (SL) to perform a geometric measurement of $H_0$ \citep{tdcosmography}.
TDC entails a measurement of this time delay and modeling of the full strong lensing system. 
$H_0$ is then inversely proportional to the time delay between photons.
This technique has been utilized with quasars (persistent variable objects) to measure $H_0$ to better than five percent precision in time-varying SL systems \citep{strides-dsplq, holicow}.

The cosmic expansion rate today $H_0$ is a critical parameter for understanding the evolution of the Universe.
There are multiple probes of $H_0$, including the cosmic distance ladder, extrapolation from the cosmic microwave background (CMB), and strongly lensed variable sources, like supernovae and quasars.
There is currently a significant tension amongst the probes \citep{wendy}, particularly between early-Universe probes like the CMB \citep{cmb} and late-Universe probes like the type-Ia SNe distance ladder anchored with Cepheid variable stars \citep{riess}.
Time-delay cosmography, as probe of $H_0$, does not rely on anchoring measurements to late-Universe objects or extrapolating from early-Universe physics, so it offers a new perspective on the expansion rate of the Universe today.

Aside from quasars, supernovae (SNe) are the other major class of common time-varying sources that are bright enough to be detected at cosmological distances.
SNe present an experimental challenge in TDC because they become bright and fade on the scale of months, therefore requiring rapid identification and analysis to obtain a measurement of the time delay.
However, the larger variability on shorter timescales compared to quasars offers a competitive advantage in measuring the time delay. 
Another advantage of LSNe is that a common subclass of SNe (SNe-Ia) can facilitate highly accurate modeling of the lensing gravitational potential -- and therefore a more precise $H_0$ measurement \citep{simon1, simon2, simon3, simon4} -- as a result of their standardizable brightness \citep{sniabrightness}.
To date, only a handful of LSNe have been detected \citep{kelly, Rodney2021, unresolved_lsn_1, unresolved_lsn_2}, and only two LSNe-Ia have been discovered \citep{iptf16geu, unresolved_lsn_3}.

Large optical surveys are ideal datasets to search for LSNe, since high area coverage and returning to the same field multiple times increase the chances of observing a LSN, though the rarity of LSNe still makes their detection a challenging problem.
For example, based on the area covered, imaging depth, and length of observations, only 0.5-2 LSNe are expected to be in DES data \citep{oguri}.
Looking forward to the next era of optical survey astronomy, the Vera C. Rubin Observatory's Legacy Survey of Space and Time \citep[LSST;][]{lsst} plans to cover the entire southern sky to greater depth and with higher cadence than any predecessor survey --- e.g. the Dark Energy Survey \citep[DES;][]{desfinale}.
Preliminary forecasts indicate that the Rubin Observatory will  detect hundreds to thousands of these systems \citep[]{goldstein, oguri, wojtak}.
Each detected time-varying SL system has the potential to produce an independent measurement of $H_0$, meaning that the measured statistical precision on $H_0$ using this technique will scale with $1/\sqrt{N}$, where $N$ is the number of detected systems.
Therefore, one of the main goals in the LSST-era is the identification and characterization of as many LSNe-Ia as possible to precisely measure $H_0$ \citep{lsstsrd}. 

Fast and robust algorithms for detecting LSNe-Ia are essential to keep pace with the data stream of LSST and surveys with comparable data size.
Furthermore, because supernovae fade after their explosion, they must be identified rapidly to facilitate follow-up observations and more detailed characterization of the system for lensing analyses.
One approach to detecting LSNe is to observe known SL systems and wait for SN.
This approach leverages existing SNe detection strategies and infrastructure and is expected to detect all SNe in known SL systems in the southern hemisphere.
However, the Rubin Observatory's LSST will probe deeper than any previous optical survey.
It is expected that a large population ($\sim10$\%) of all SL systems will have a source galaxy that is too faint to have been detected previously, and they will be missing from the list of target systems \citep{watchlist}.
Another approach leverages the standardizable brightness of SNe-Ia and proposes the search for brighter than expected (due to lensing magnification) SNe near elliptical galaxies \citep[]{goldstein}.
This magnitude threshold technique is expected to detect $\sim500$ LSNe-Ia, but is specific to elliptical lens galaxies and will miss LSNe-Ia whose date of peak brightness do not align with the LSST cadence. 

In this work, we introduce a deep neural network architecture designed to identify LSNe-Ia without the requirements of targeting known SL systems, elliptical galaxies, or observing the peak brightness of the SN.
Deep learning algorithms have been highly successful, both in terms of accuracy and speed, in the fields of image-based SL system detection \citep[among several others]{jacobs} and light-curve-based SNe classification \citep[among several others]{supernnova}.
A convolutional neural network \citep[CNN;][]{cnn} is a kind of deep algorithm that slides learnable matrix operators along images to emphasize or de-emphasize characteristic shapes, such as lensing arcs.
Recurrent neural networks \citep[RNNs;][]{rnn} can model sequences of data, such as lightcurves, to make classifications based on how the data vary in time.
Within the RNN architecture, long short-term memory (LSTM) networks \citep{lstm}, which introduce cells with learnable pathways for information to be passed along, have demonstrated improved performance in sequence characterization.
In our approach, we combined a CNN and an LSTM as two branches and optimally utilize the information from both branches --- an architecture that leverages the data structures of each input and is tailored to the challenge of immediate LSNe-Ia identification. 
Other studies, such as \citet{spaciotemporal}, have combined convolutional and recurrent layers in different deep learning architectures to target LSNe identification, though our multi-branch approach is unique in that it places the spatial and temporal information on the same footing from the start.

We present this work as follows:
In Section \ref{sec:methods}, we describe the simulations used for training and testing \ZipperNet, the data processing procedure for utilizing both spatial and time series information, and the architecture of \ZipperNet. 
In Section \ref{sec:results}, we describe the results from applying \ZipperNet\ to four simulated datasets that emulate modern survey data products.
In Section \ref{sec:discussion}, we discuss the performance of \ZipperNet\ with respect to the dataset properties and the network architecture. 
We conclude in Section \ref{sec:conclusion}.

%% file: methods.tex
\subsection{Data Simulation}
\label{sec:data}

\begin{figure*}
    \centering
    \includegraphics[trim={0cm 2.5cm 0cm 1.5cm},clip,width=\textwidth]{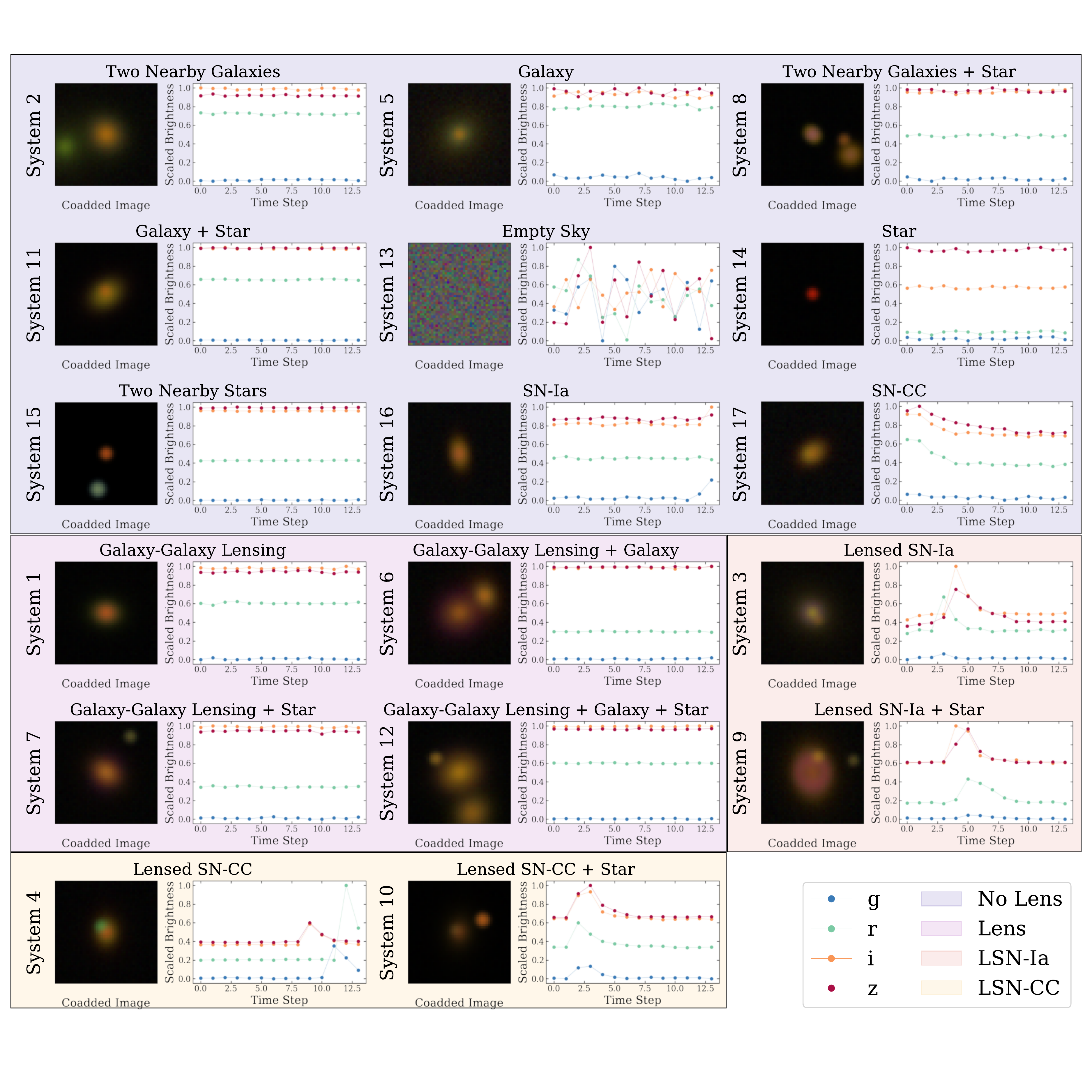}
    \caption{Examples of the 17 simulated systems from the LSST-wide dataset grouped into No Lens (blue), Lens (magenta), LSN-Ia (orange), and LSN-CC (yellow) classes. 
    Each example displays a coadded composite image of the $gri$ bands and the extracted lightcurve from our processing in Section \ref{sec:processing}. 
    Each time step in the LSST-wide dataset is approximately 12 days.
    }
    \label{fig:data}
\end{figure*}

% To train and test ZipperNet, w
We simulated images of astronomical strong lensing systems with the open-source software package, \texttt{deeplenstronomy} \citep{deeplenstronomy}, which is built around  \texttt{lenstronomy} \citep[]{lenstronomy, lenstronomy2}, a widely used package that performs gravitational lensing calculations, modeling, and simulations in a variety of contexts.
\texttt{deeplenstronomy} provides additional features that are important for the era of large-scale surveys and deep learning studies of strong lenses -- e.g., image and SN injection, probability distribution sampling, and realistic observing conditions.

\subsubsection{Survey Emulation}
We simulated four datasets, distinguished by their camera specifications, observing conditions, and cadence, each emulating a distinct modern or next-generation cosmic survey: one wide- and one deep-field dataset for DES -- DES-wide and DES-deep, respectively; a wide-field LSST dataset (LSST-wide), and a three-day cadence Dark Energy Camera \citep[DECam;][]{brenna} dataset similar to the Dark Energy Spectroscopic Instrument DECam Observation of Transients (DESI-DOT) program (Palmese \& Wang; DECam Proposal 2021A-0148).
All datasets utilize the $g$, $r$, $i$, and $z$ optical filters and include 45-by-45-pixel images.
% , and are summarized in Table \ref{tab:datasets}.
The DES-wide, DES-deep, and DESI-DOT datasets simulate images from the DECam, which connects the pixel size, gain, and read noise across those datasets.
The LSST-wide dataset simulates LSSTCam \citep{lsstcam} images, which have similar but slightly adjusted values for the camera properties.

% :  DES wide field (DES-wide), DES deep field  (DES-deep), LSST wide field  (LSST-wide), and a three-day cadence DECam survey similar to the Dark Energy Spectroscopic Instrument DECam Observation of Transients (DESI-DOT) program (Palmese \& Wang; DECam Proposal 2021A-0148).

% The datasets are .

The DES-wide and DESI-DOT datasets both use the real observing conditions (seeing and sky brightness) from the DES wide-field survey \citep{desdr1}.
For DES-wide and DESI-DOT, the exposure times are 90 and 60 seconds, respectively.
The DES-deep dataset has different seeing, sky brightness, zeropoint, and exposure times chosen for the DES SN program \citep[]{dessn}.
In general, these exposure times are on the order of 200 seconds, but the acceptable seeing criteria can be worse than the DES wide-field survey.
The LSST-wide observing conditions are estimated from simulations of the first year of the survey and utilize 30-second exposures \citep[]{lsst_table}.

Each dataset has a specific and distinct cadence, and the density of observations significantly affects the analysis in this work.
The LSST-wide, DESI-DOT, and DES-deep datasets contain a baseline of 14 epochs per band.
While the LSST main survey cadence is still being designed at the time of this writing, our fiducial 14-epoch data sequences are sufficiently short that they are obtainable from both the ``baseline'' and ``rolling'' cadences that are under consideration for the survey.
The DES-wide dataset contains seven exposures in each band spread over 5.5 years to match the real survey.
We sampled the observation times of several fields from the DES footprint to generate the \texttt{deeplenstronomy} simulations.
The LSST-wide cadence is estimated using several realizations of an intra-band spacing of $12\pm5$ days over a three-month period \citep[]{lsst_table}.
The DES-deep cadence is estimated using several realizations of an intra-band spacing of $6\pm1$ days over a one-month period \citep[]{dessn}.
Lastly, the DESI-DOT cadence is an exposure in each band every three nights over a one-month period.

We seek to avoid jargon confusion between astronomy and machine learning contexts with respect to the term, ``epoch.'' 
In this work, ``epoch'' refers to one astronomical exposure or data-collection period.
When discussing neural network training steps, we use the term ``training iteration'' instead of the traditional ``epoch'' that is used in machine learning.
The data sets are summarized in Table \ref{tab:datasets}.
All \texttt{deeplenstronomy} input files for this analysis are accessible for reproduction of the datasets in \citet{zenodo}.

\subsubsection{Object, System, and Population Simulation}

% The systems simulated are the same across all four datasets.
In total, we simulate 17 different types of astronomical systems to reflect the diversity of systems that classifiers are likely to encounter when applied to observed optical survey data: (1) one galaxy behind one foreground galaxy; 
(2) two galaxies at the same redshift and with small angular separation (1 to 4 arcseconds); 
(3) one galaxy (with one SN-Ia) behind one foreground galaxy;
(4) one galaxy (with one SN-CC) behind one foreground galaxy;
(5) one galaxy;
(6) two galaxies (with small angular separation and at the same redshift) in front of one background galaxy that is at a higher redshift;
(7) one galaxy behind one foreground galaxy that has one star from the Milky Way in the image cutout;
(8) two galaxies with small angular separation and at similar redshifts that has one star from the Milky Way in the image cutout;
(9) one galaxy (with one SN-Ia) behind one foreground galaxy that has one star from the Milky Way in the image cutout;
(10) one galaxy (with one SN-CC) behind a foreground galaxy that has one star from the Milky Way in the image cutout;
(11) one galaxy that has one star from the Milky Way in the image cutout;
(12) two galaxies with small angular separation and at similar redshifts in front of one background galaxy at a higher redshift and one star from the Milky Way in the image cutout; (13) empty sky;
(14) one Milky Way star;
(15) two Milky Way stars;
(16) one galaxy with one SN-Ia; and
(17) one galaxy with one SN-CC.
Figure \ref{fig:data} shows sample images and time series  of the 17 systems.

To further enhance realism, the properties of all simulated objects are drawn from real data: all inherent physical correlations of these parameters are included in our dataset.
First, a galaxy that enters the simulations as the lens has properties drawn from a population of $\sim2,000$ observed galaxies.
The velocity dispersion and spectroscopic redshift were measured by the Sloan Digital Sky Survey \citep[SDSS;][]{sdss}.
We obtain a color-independent ellipticity, as well as a band-wise half-light radius, Sersic profile index, magnitude from 
DES Year 1 data \citep[]{morphology, desdr1}. 
Next, a putative source galaxy, whether used in a lensing or non-lensing system, draws its properties from a population of $\sim500,000$ galaxies measured by DES \citep{desdr2}.
The band-wise magnitudes of foreground Milky Way stars  were also drawn from DES data \citep{desdr2}.
Finally, the SNe were injected using public SN spectral energy distributions \citep[]{sntemplates} available in \texttt{deeplenstronomy}, which redshifts the distribution and calculates the observed magnitude in each band.
The injected SN reaches peak brightness anytime between 20 days before the first observation and 20 days after the final observation, so the dataset contains falling lightcurves, rising lightcurves, and complete lightcurves.
We do not include the effects of microlensing in our simulated dataset, because it is expected to be small compared to the change in brightness observed from a SN \citep{microlensing}.

% and DES: velocity dispersion, spectroscopic redshift, 
% ellipticity, along with half-light radius, Sersic profile index, and magnitude all on a band-by-band basis. 

% Galaxy light profiles are modeled by a Sersic profile.
% The velocity dispersion and spectroscopic redshift were measured by SDSS while the other properties were measured in DES year 1 data \citep[]{morphology, desdr1}. 

For all four survey emulation datasets, we simulate the same strong lensing systems: all strong lensing systems are emulated in all four of the cosmic survey contexts.
While our simulated datasets are subject to selection biases from the detection limits of DES and SDSS for source and galaxies, respectively, the data nonetheless contain realistic collections of object properties, which enables the validation of our deep learning detection method on realistic survey data.

\subsection{Data Processing}
\label{sec:processing}

\texttt{deeplenstronomy} emulates observational surveys by producing a time series of images.
With the exception of small effects from observing conditions, images in a series will be approximately identical, because astronomical objects are approximately stationary on month-long timescales. 
Even in the case of a SN, the primary difference is the presence of one or more point sources in some of the images in the time series.
We condensed the image information to single-image input for \ZipperNet\ by averaging all images in the time series on a pixel-by-pixel basis within each band.
This processing reduces noise fluctuations from the observing conditions while preserving the presence of SNe, thus increasing the overall signal-to-noise ratio of the image and making faint objects more visible.
After averaging the images, the pixel values of the mean images are scaled to range from 0 to 1 on a per-example basis to preserve color relationships.

To concisely characterize the temporal behavior of a time series of images and to avoid relying on source identification or deblending algorithms, we follow a process that  reflects a standardized background-subtracted aperture flux measurement in astronomy.
We measure the signal (S) and background (B) to extract a background-subtracted brightness (S $-$ B) of individual images within a predefined circular aperture at the center of each image.
In equation form, this process can be expressed as
\begin{align}
    &\textrm{S} = \sum^N_{i,j}W_{i,j}X_{i,j}, \label{eq:signal}\\
    &\textrm{B} = \underset{W_{i,j} = 0}{\textrm{median}}\left[ X_{i,j}\right] \times \sum^N_{i,j}W_{i,j},\label{eq:background}\\
    &\textrm{Brightness} = \textrm{S} - \textrm{B},
    \label{eq:brightness}
\end{align}
where $i$ and $j$ index the row and column of the image pixels, $N$ gives the number of pixels along one dimension of the images, $X$ is an image, and $W$ is the aperture.
$W_{i,j}$ is zero outside the aperture and one inside the aperture.
In this work, the circular aperture has a 20-pixel radius, which corresponds to 5.26 arcseconds for DECam and 4.0 arcseconds for LSST-Cam -- both much larger than typical galaxy-scale lens Einstein radii, which are approximately in the range [0.5, 1.2] arcsec.
For processing within the neural networks, we again scale the extracted brightness to between 0 and 1 on a per-example basis.

A byproduct of this process is the significant increase in noise in the photometry measurements;  we find, however, that this effect does not hinder the deep learning methods.
After the averaging images and extracting lightcurves, each example input to \ZipperNet\ is a 45-pixel-by-45 -pixel image in each band and a lightcurve of the extracted brightness at each time step in each band.
The operations required to extract the photometry of the systems have little computational cost and can be easily broadcasted, which is a great benefit considering the scale of modern astronomical datasets.

Lastly, we define a classification scheme for our 17 simulated systems.
We construct a four-class problem, where the classes are ``No Lens,'' ``Lens,'' ``LSNe-Ia,'' and ``LSNe-CC,'' as shown in Figure \ref{fig:data}.
The No Lens class collects the cases where there is no gravitational lensing present -- labeled as 2, 5, 8, 11, 13, 14, 15, 16, and 17.
The Lens class collects cases where there is gravitational lensing but no SNe in the background galaxy -- labeled as 1, 6, 7, and 12.
The LSNe-Ia  and LSNe-CC classes collect cases with gravitational lensing and a SN in the background galaxy -- labeled as 3 and 9 and as  4 and 10, respectively.
Each of the four classes contain 1,250 examples, with equal representation of the individual constituent cases.
To augment the datasets and increase the size of each class eightfold to 10,000, we rotated and mirrored the images; the lightcurve was unaffected due to the circular aperture extraction method.
When we split the datasets into smaller training and testing datasets, none of the examples in the testing dataset are rotated or mirrored versions of objects in the training dataset: the two datasets are rotated and mirrored independently.

\subsection{ZipperNet}
\label{sec:network}

\begin{figure*}
\centering
\includegraphics[trim={0cm 1.7cm 0cm 0cm},clip,width=0.95\textwidth]{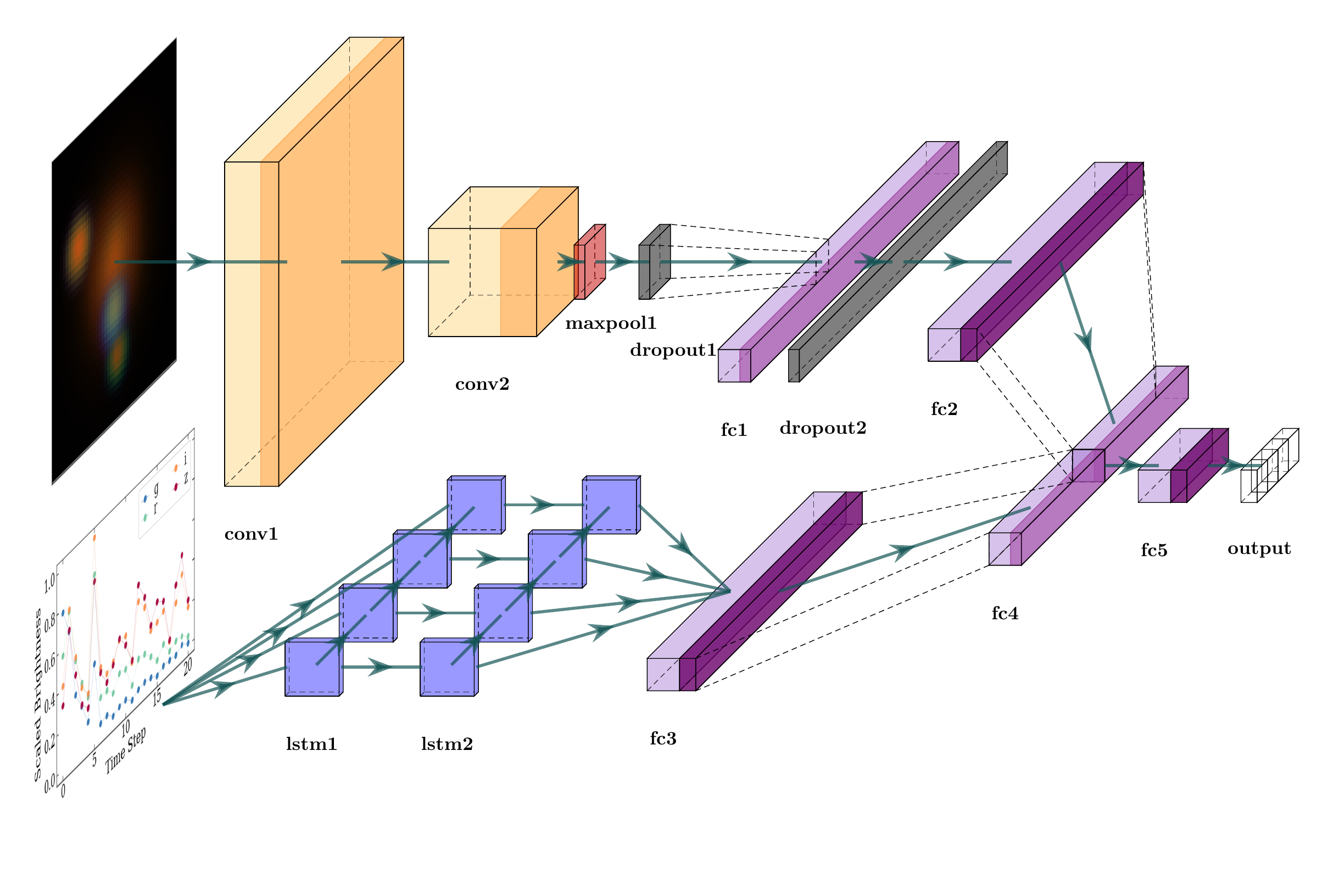}
\caption{A diagram of the ZipperNet architecture. 
Two convolutional layers (orange) receive $griz$ images while two LSTM layers (blue) receive extracted $griz$ lightcurves. 
Fully-connected layers (purple) process the flattened and concatenated outputs. 
Nonlinear activation functions are indicated by a darkened band at the right edge of a layer. 
Table \ref{tab:network_params} displays the layer specifications. 
The final output is a length-4 array, with each element representing the score of one of the four classes in the dataset.
This visualization was made with the \texttt{PlotNeuralNet} library \citep[]{network_vis}.}
\label{fig:network}
\end{figure*}

Our \ZipperNet\ architecture is designed to treat the image-based information and the lightcurve-based information on equal footing.
On one branch, the images are passed through convolutional layers, flattened into one dimensional arrays, and condensed in size.
On another branch, the lightcurves are passed through recurrent layers composed of LSTM cells and flattened into one dimensional arrays.
The flattened images and lightcurves output by each branch are condensed to equal sizes, concatenated, and then mapped to four output features --- one for each class of our problem.
We then obtain a single classification by determining which of the four output features has the largest value.
By zipping convolutional layers and recurrent layers into one coherent deep learning architecture (in joining the branches), the training of the network will optimize weights in both types of layers simultaneously.
The architecture is illustrated in Figure \ref{fig:network}, and the specifications of each layer are presented in Table \ref{tab:network_params}.
All deep learning code in this analysis utilizes the \texttt{PyTorch} \citep[]{pytorch} library.

For each of the four datasets in our analysis, we trained an individual \ZipperNet: we trained on 90\% (9,000 samples) of the simulated data and used the remaining 10\% (1,000 samples) for testing.
We did not utilize any of our simulated data as a validation dataset for hyperparameter optimization.
Rather, we chose the ZipperNet hyperparameter settings based on an independent toy dataset composed of images of different shapes (squares versus circles) with different time-varying properties (parabolic versus linear change in total brightness) and fixed the settings for each of the four ZipperNet instances.
This choice is motivated by a desire to keep the model constant and prevent the hyperparameter settings from favoring one of the simulated datasets over another.
We therefore control confounding variables in our experiment such that we can connect differences in model performance between the four simulated datasets to dataset properties.

We chose a batch size of five for the training because LSTM cells generally perform better when processing smaller amounts of information at the same time.
We also utilized categorical cross entropy as the loss function of the network and a learning rate of 0.001 with the Adam \citep{adam} optimizer.
As the network trained, we monitored the accuracy (the number of correct classifications divided by the total number of samples) for the training and testing datasets.
In each case, the training and testing accuracy plateaued after $\sim10$ training iterations, but we allowed the training to continue for 40 training iterations.
The fully trained network is chosen as the point during training with the highest testing accuracy.
We use the term ``training iteration'' in place of the traditional ``epoch'' to avoid confusion with the astronomy term ``epoch'' utilized in other parts of this analysis.
The accuracy for the training and testing sets for each of the four datasets is presented in Table \ref{tab:accuracy}.

\begin{table}
    \centering
    \begin{tabular}{|c|l|} \toprule
     Layer    &  Specifications \\ \midrule
     \textbf{conv1}$^\dagger$   &  Conv2D --- ($k$: 15, $p$: 2, $s$: 3) --- (4 $\rightarrow$ 48) \\ 
     \textbf{conv2}$^\dagger$   &  Conv2D --- ($k$: 5, $p$: 2, $s$: 1) --- (48 $\rightarrow$ 96) \\ 
     \textbf{maxpool1}   & MaxPool2D ($k$: 2) \\ 
     \textbf{dropout1}   & Dropout2D ($f$: 0.25) \\ 
     \textbf{fc1}$^\dagger$   & Fully-Connected (3456 $\rightarrow$ 408) \\ 
     \textbf{dropout2}   & Dropout ($f$: 0.5) \\ 
     \textbf{fc2}$^\ddagger$   & Fully-Connected (408 $\rightarrow$ 25) \\ \midrule
     \textbf{lstm1}  & LSTM ($h$: 128) \\
     \textbf{lstm2}  &  LSTM ($h$: 128) \\
     \textbf{fc3}$^\ddagger$   & Fully-Connected (128 $\rightarrow$ 25) \\ \midrule
     \textbf{fc4}$^\dagger$   & Fully-Connected (50 $\rightarrow$ 8) \\ 
     \textbf{fc5}$^\ddagger$   & Fully-Connected (8 $\rightarrow$ 4) \\  \bottomrule
    \end{tabular}
    \caption{\ZipperNet\ layer specifications. 
    We use the following shorthand: kernel size ($k$), padding ($p$), stride ($s$), dropout fraction ($f$), and hidden units ($h$). 
    Arrows indicate the change in the size of the data representation as it is passed through the layer.  ``$^\dagger$'' indicates a Rectified Linear Unit (ReLU) activation function. ``$^\ddagger$'' indicates a LogSoftmax activation function. In total, our model contains 1,551,632 trainable parameters.}
    \label{tab:network_params}
\end{table}

\begin{figure*}
    \centering
    \includegraphics[trim={2cm 0cm 0.5cm 0cm},clip,width=0.48\textwidth]{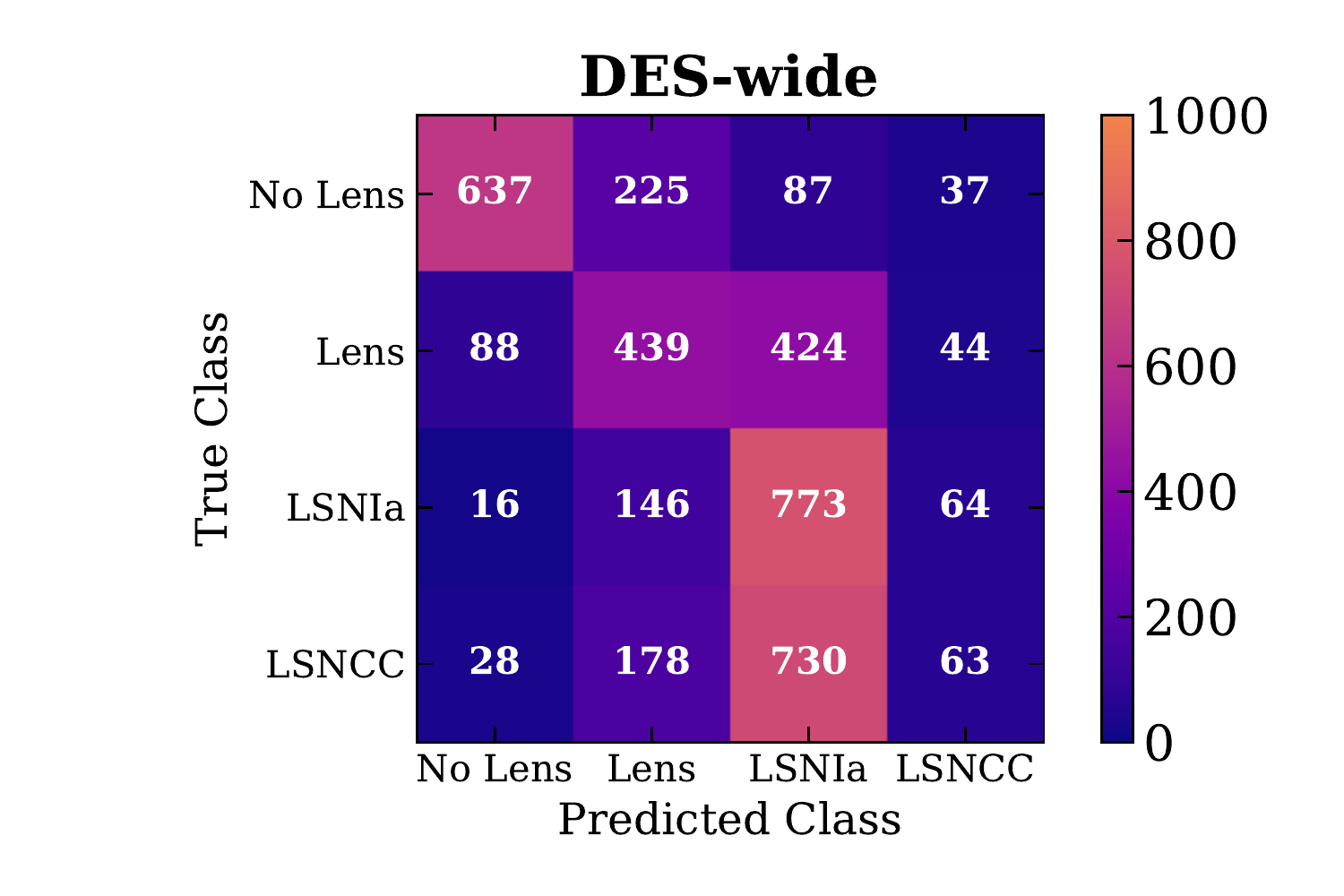}
    \includegraphics[trim={2cm 0cm 0.5cm 0cm},clip,width=0.48\textwidth]{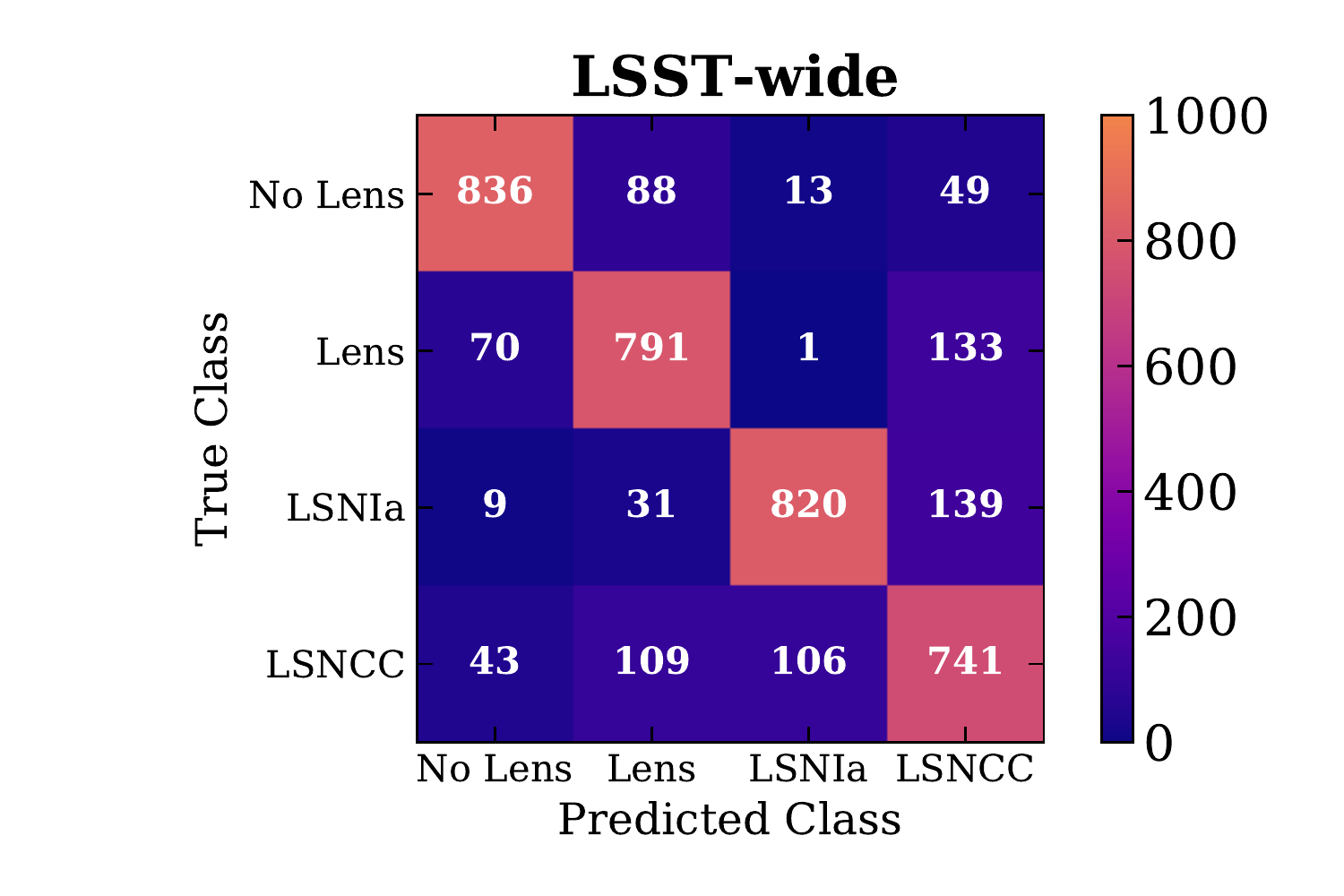}
    \includegraphics[trim={2cm 0cm 0.5cm 0cm},clip,width=0.48\textwidth]{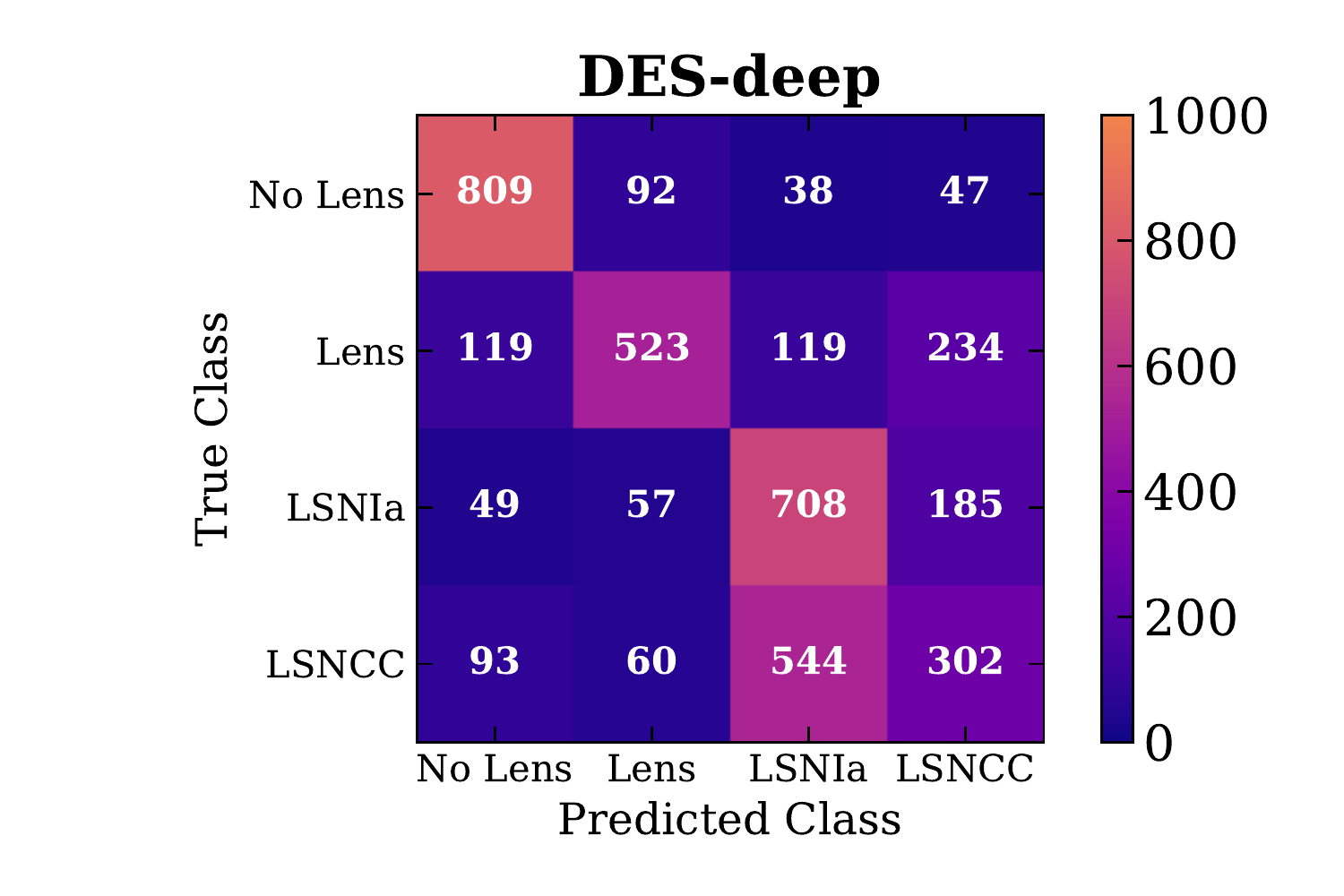}
    \includegraphics[trim={2cm 0cm 0.5cm 0cm},clip,width=0.48\textwidth]{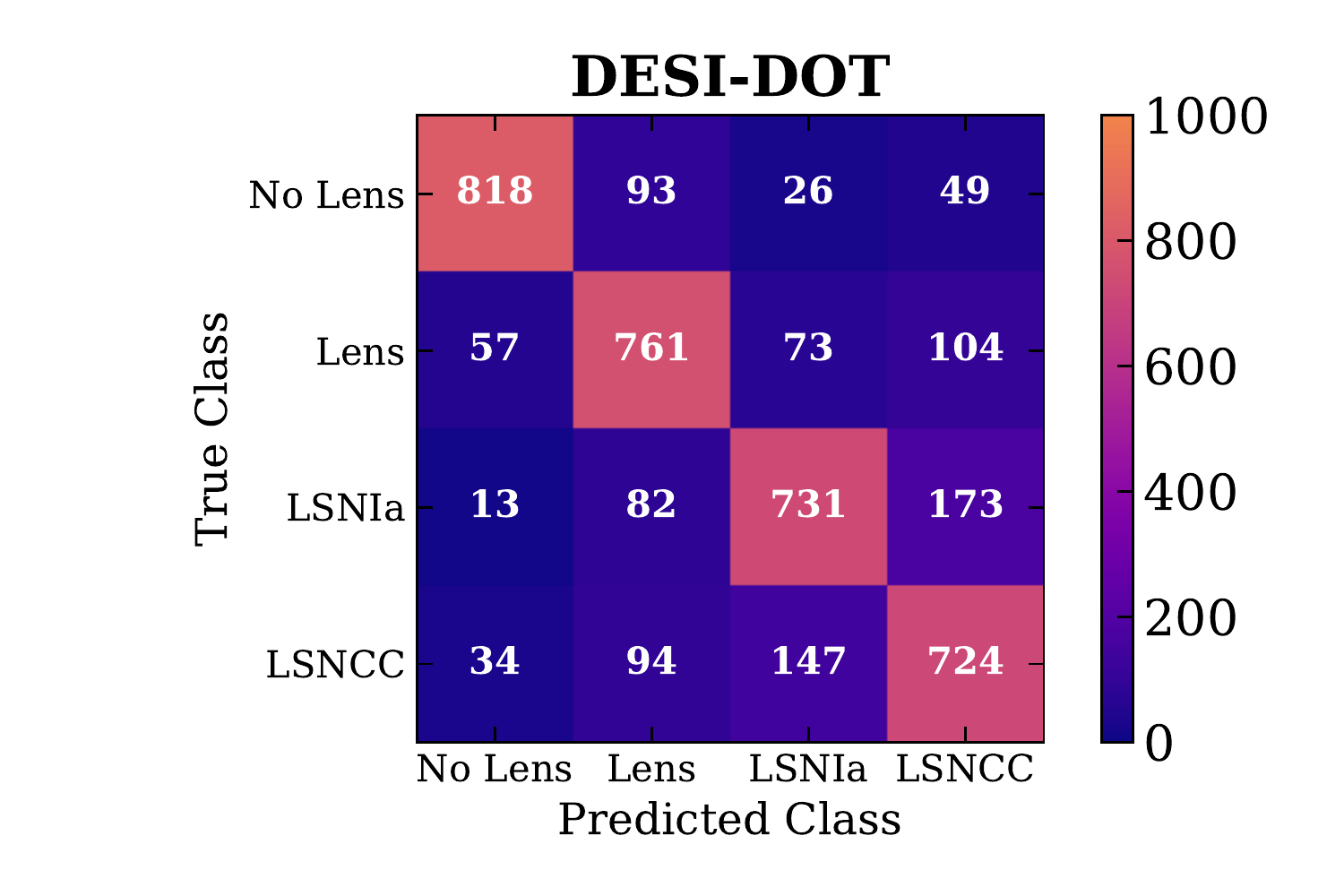}
    \caption{Confusion matrices from applying a trained \ZipperNet\ to the DES-wide, LSST-wide, DES-deep, and DESI-DOT datasets. 
    The matrices display the predictions of \ZipperNet\ as a function of the true classes of the examples for the test datasets. Each row in each matrix has 1,000 examples in total, so the matrix counts are representative of normalized percentages.}
    \label{fig:confusion}
\end{figure*}

%% file: results.tex
In general, we find that the \ZipperNet\ model is capable of identifying SL systems, identifying the presence of an SN within SL systems, and distinguishing between LSN-Ia LSN-CC.
Furthermore, we demonstrate that \ZipperNet\ can perform these classifications on simulated datasets across wide ranges of depth, observing conditions, and cadence.
We use a four-class confusion matrix to compare the predicted and true labels resulting from \ZipperNet's execution (see Figure \ref{fig:confusion}).
In all the matrices, the strong representation along the main diagonal indicates correct classifications.
ZipperNet had the lowest performance on the DES-wide data, and we discuss this result in Section \ref{sec:discussion}.
% for each dataset is displayed in
% , where the predictions made by ZipperNet in our four class scheme (No Lens versus Lens versus LSNe-Ia versus LSNe-CC) are compared to the true classes of the examples.

% The main sources of confusion are straightforward to interpret.
There are multiple primary sources of confusion.
% We note confusion between the No Lens class and Lens class, as well as confusion between the LSNe-CC class and the LSNe-Ia class.
Confusion between the No Lens and Lens classes is likely due to pixel-based features being difficult to distinguish from the images.
This confusion is the strongest within the DES-wide dataset, so we attribute this behavior to the optical depth of the images, since the DES-wide dataset is the shallowest dataset simulated and faint source galaxies would become more difficult to identify.
In the deeper datasets, this confusion is caused by exposures with high seeing or systems with small Einstein radii, where in both cases objects blur together.
% The second source of confusion is between the LSNe-Ia and LSNe-CC classes.
The confusion between the LSNe-Ia and LSNe-CC classes is likely due to the difference in cadence and seeing in the surveys.
DES-deep, LSST-wide, and DESI-DOT are much higher cadence datasets than DES-wide (See Table \ref{tab:datasets}), indicating that more densely sampled SN lightcurves are easier for the LSTM cells to classify.
A comparison of the DES-deep and DES-wide confusion matrices indicates that there are situations where cadence can be more important than seeing (specifically noting the LSNe-Ia versus LSNe-CC confusion), since the DES-wide dataset had better seeing than DES-deep, but that these situations require dramatic differences in cadence density.
Furthermore, the LSST-wide and DESI-DOT datasets have much better seeing than the DES-deep dataset, which shows the importance of being able to resolve spatial features when making classifications.
We discuss the importance of the cadence and seeing in more detail in Section \ref{sec:discussion}.

\begin{table}
    \centering
    \bigskip
    \begin{tabular}{|c|cc|cc|}
    \toprule
      & \multicolumn{2}{|c|}{Training Acc.} & \multicolumn{2}{|c|}{Testing Acc.} \\ \midrule
      Dataset  & Iter. 10 & Iter. 40 & Iter. 10 & Iter. 40 \\ \midrule
      \textbf{DES-wide} & 0.496 & 0.503 & 0.473 & 0.487 \\
      \textbf{LSST-wide} & 0.750 & 0.838 & 0.709 & 0.785 \\
      \textbf{DES-deep} & 0.550 & 0.613 & 0.535 & 0.573 \\
      \textbf{DESI-DOT} & 0.737 & 0.813 & 0.680 & 0.735 \\
    \bottomrule
    \end{tabular}
    \caption{Training and testing accuracy for \ZipperNet\ on each of the four datasets (rows) at training iterations 10 and 40 (columns).}
    \label{tab:accuracy}
\end{table}

\begin{figure}
    \centering
    \includegraphics[trim={0.5cm 0.5cm 0cm 0cm},clip,width=0.47\textwidth]{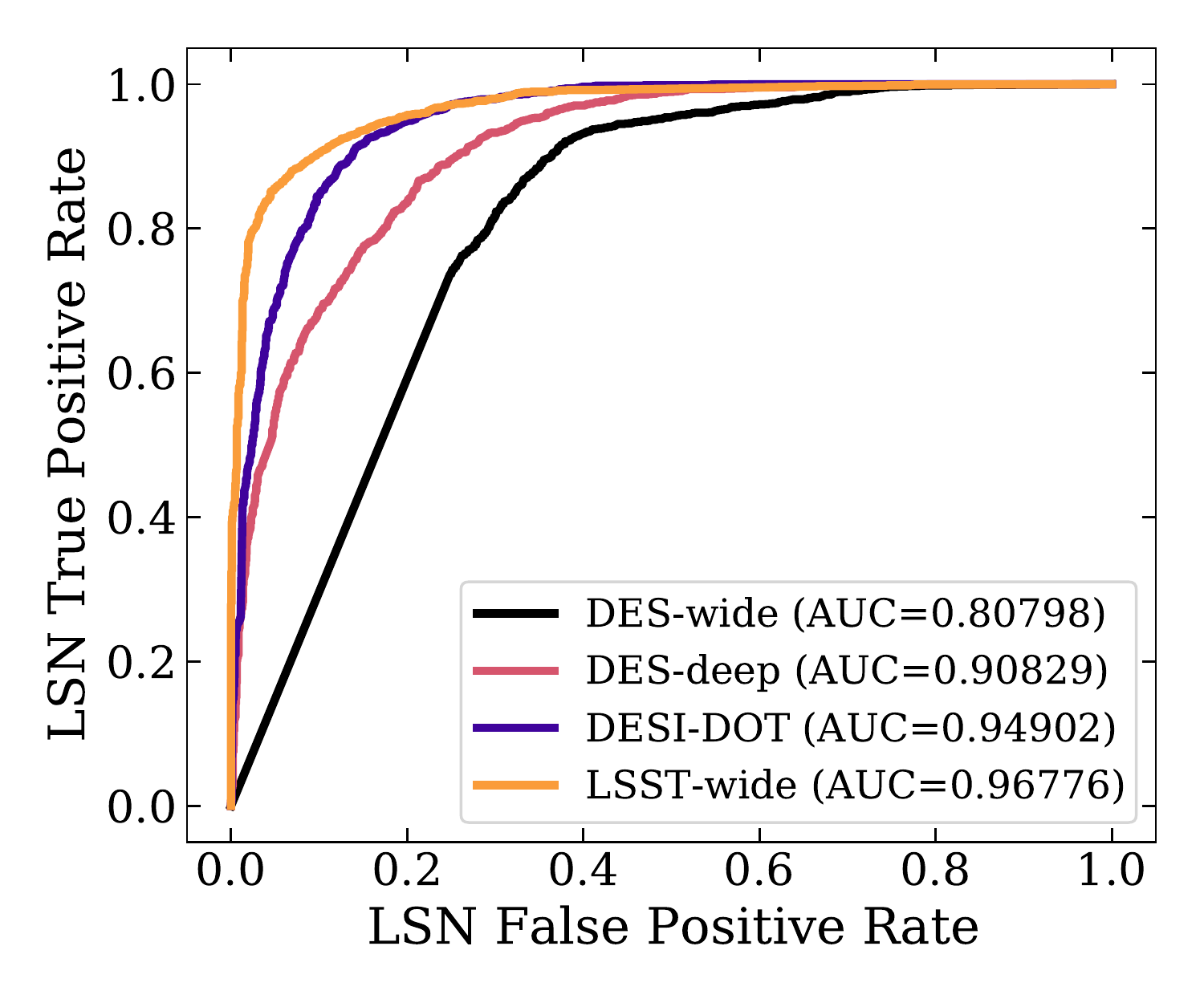}
    \caption{Receiver Operating Characteristic (ROC) curves for the LSST-wide, DES-deep, and DESI-DOT datasets. 
    The ROC curves are calculated for the two-class problem of LSNIa and LSNCC versus No Lens and Lens. 
    An Area Under Curve (AUC) of 1.0 indicates perfect performance, while an AUC of 0.5 indicates random guessing.}
    \label{fig:roc}
\end{figure}

In practice, a general LSN identifier is itself a useful tool: LSNe-CC can be utilized for time-delay cosmography measurements, though they offer less precision on the final $H_0$ measurement.
If we reframe this classification scheme from a four-class problem to a two-class problem (No Lens and Lens in one class and LSNe-Ia and LSNe-CC in second class), the performance is boosted.
We can obtain two-class problem classifications from our four-class network outputs by selecting the class predicted by the network and sorting it into LSN or not LSN.
Figure \ref{fig:roc} shows a Receiver Operating Characteristic (ROC) curve for the DES-wide, LSST-wide, DES-deep, and DESI-DOT datasets.
ROC curves are standard tools for assessing the predictive power of a classifier by calculating the false positive rate and true positive rate at all possible probabilities output by the classifier.
A perfect classifier will have an Area Under Curve (AUC) of 1.0 while a classifier that guesses randomly will have an AUC of 0.5.
\ZipperNet\ shows high performance when classifying LSNe versus everything else, and this high performance extends across the LSST-wide, DES-deep, and DESI-DOT datasets.
This result can also be interpreted from the confusion matrix (Figure \ref{fig:confusion}), where there is little confusion between the LSNe classes and non-LSNe classes.
We comment on this high performance in the context of ZipperNet's architecture in Section \ref{sec:discussion}.

We also estimate the baseline true-positive rate and false-positive rate for LSNe using this technique in the different datasets.
When applying the \ZipperNet\ technique to real data, we would expect real data to be used in the training and validation, which would produce more accurate estimates.
That being said, we can initially report a LSN true (false) positive rate of 90.2~\% (29.6~\%) for DES-wide, 87.0~\% (21.9~\%) for DES-deep, 91.5~\% (9.8~\%) for LSST-wide, and 89.0~\% (12.6~\%) for DESI-DOT.
The true-positive rate of approximately 91.5~\% for LSST-wide is the higher than the estimated recovery rates of LSNe for the non-deep-learning approaches mentioned in Section \ref{sec:introduction}.
Furthermore, with relatively low false-positive rates, we do not expect the data stream of the LSST to be overwhelmed by other astronomical systems being incorrectly labeled as LSNe.

In the two-class problem, we performed additional analysis to interpret the features \ZipperNet\ identified for making classifications.
Figure \ref{fig:gallery} displays examples from the LSST-wide dataset arranged into groups of correctly classified LSNe (true positives), other astronomical systems correctly classified as not LSNe (true negatives), other astronomical systems erroneously labeled as LSNe (false positives), and LSNe that were missed (false negatives).
These examples were found to be representative of the general relationship between the properties of objects and the predictions made by \ZipperNet.
\ZipperNet\ is able to correctly classify LSNe when the Einstein radius is small ($0.5 - 1.0$ arcsec) and with foreground stars in the image, both of which would trouble a standalone CNN.
Upon close inspection, the true-positive images display galaxies with non-uniform light profiles, hinting at the presence of lensing, but the dominating feature is clearly the large fluctuation in brightness detected by the LSTM.
The true negatives show systems with evidence of lensing as well, but this time there is no temporal behavior to indicate the presence of a SN.
The false positives contain images of lensing or crowded fields with small, but non-negligible coherent time-varying behavior; they in general contain both spatial and temporal features similar to the LSNe class.
Lastly, the false negatives are the most important group to understand due to the rareness of LSNe.
In some cases, the presence of a star in the aperture used to extract the scaled brightnesses can be bright enough to obscure the change in brightness of the LSNe.
Similarly, if the source galaxy is distant and the alignment of the lensing system does not produce sufficient magnification, the imaging may not be deep enough to see a lensed source or a background SNe.
Both these cases of false negatives demonstrate difficult to detect systems with LSNe and point toward a physically-motivated selection as opposed to inaccurate feature representations learned by \ZipperNet.
In general, we find that \ZipperNet\ learns features we would \textit{a priori} expect.

%% file: discussion.tex
\begin{figure*}
    \centering
    \includegraphics[trim={0cm 2cm 0cm 1.5cm},clip,width=\textwidth]{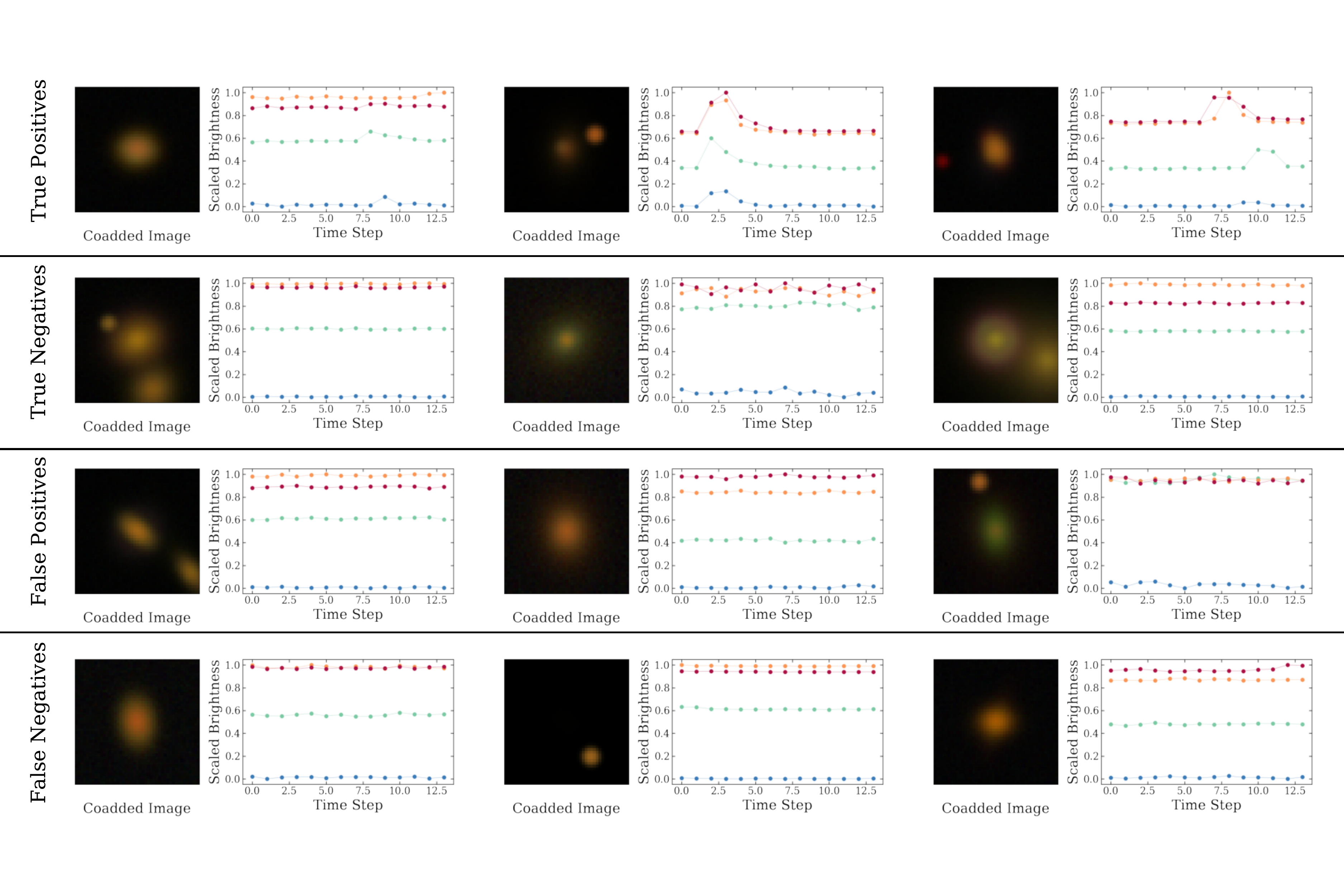}
    \caption{Examples of classifications made by ZipperNet from the LSST-wide dataset, grouped by true positives, true negatives, false positives, and false negatives. The color scheme for the bands used in the lightcurves is the same as Figure \ref{fig:data}.}
    \label{fig:gallery}
\end{figure*}

In general, \ZipperNet\ can identify SL systems, identify LSNe within those systems, and classify the LSNe as LSNe-Ia or LSNe-CC.
The variance in the performance across different datasets indicates a correspondence between data quality and LSN identification power.
% aspects of requisite data quality for LSNe identification.
% First, the observational seeing  of data quality on predictive power is the seeing of the observations.
The DES-deep and DESI-DOT datasets have slightly higher cadences (more samples within a time series) than the LSST-wide dataset, but the \ZipperNet\ accuracy was considerably higher for the LSST-wide dataset, which has higher seeing (lower image quality). 
The DES-deep dataset emulates that of the DES SNIa observing program, in which exposures were collected on nights with slightly poorer (higher) seeing to optimize the seeing of the exposures used for weak lensing measurements.
In DES data processing and SNIa analysis, the difference-imaging \citep[]{diffimg} and scene-modeling \citep[]{smp} techniques can nevertheless detect and measure SNe when seeing is up to 2~$\arcsec$.
We bypassed those time-consuming techniques with our circular aperture extraction method for the lightcurve brightnesses.
Without those techniques, the performance of \ZipperNet\ is degraded, because the higher seeing in the DES-deep dataset obscures image patterns that would otherwise be detectable in LSST-wide and DESI-DOT datasets.

% -- which we avoid with our circular aperture extraction method --

% Thus, in the context of the larger DES, the SNIa-program exposures were collected on nights with slightly poorer seeing to optimize the seeing of the exposures used for weak lensing measurements.
% The larger seeing in the DES-deep dataset obscures the image patterns ZipperNet was able to detect in the LSST-wide and DESI-DOT datasets.

%%%%%%%%%%%%%%%%%%%%%%%%%%%%%%%%%%%%%%%%%%%%%%%%%%%%%%%
%%%%%%%%%%%%%%%%%%%%%%%%%%%%%%%%%%%%%%%%%%%%%%%%%%%%%%%

A second data quality factor in \ZipperNet's predictive power is the cadence of the observations.
The DES-wide dataset has excellent depth and seeing, but a low-density time sampling: \ZipperNet\ fails to perform at similar levels to the other datasets.
With DESI-DOT, which has high cadence and similar depth and seeing as DES-wide, \ZipperNet\ was able to learn the underlying features of the four classes extremely well.
Overall, we find that together seeing of $\lesssim 1.2$~$\arcsec$ (corresponding to typical upper limits on Einstein radii of galaxy-scale lenses) and a cadence with intra-band spacing of $\lesssim15$~nights  (corresponding to roughly the timescale for SNe evolution) can also improve performance.

Nevertheless, even when \ZipperNet\ had confusion between LSN-Ia and LSN-CC (primarily caused by deficient cadence of the dataset) or confusion between the No Lens and Lens classes (primarily caused by seeing greater than typical Einstein radii), \ZipperNet\ performs extremely well as a LSNe finder.
Reducing the classification to two classes -- LSN and everything else -- shows high performance for the LSST-wide, DES-deep, and DESI-DOT datasets (Figure \ref{fig:roc}).
The \ZipperNet\ architecture as a LSNe finder in this setting does not suffer from the dependence on seeing observed in the four-class problem and is slightly less dependent on the cadence.
By balancing the image and temporal inputs, ZipperNet finds weightings and combinations of the two data products optimal for LSNe detection.
We interpret this result as a demonstration of a key strength of the \ZipperNet\ architecture.

\begin{figure}
    \centering
    \includegraphics[trim={0.5cm 0.5cm 0cm 0cm},clip,width=0.47\textwidth]{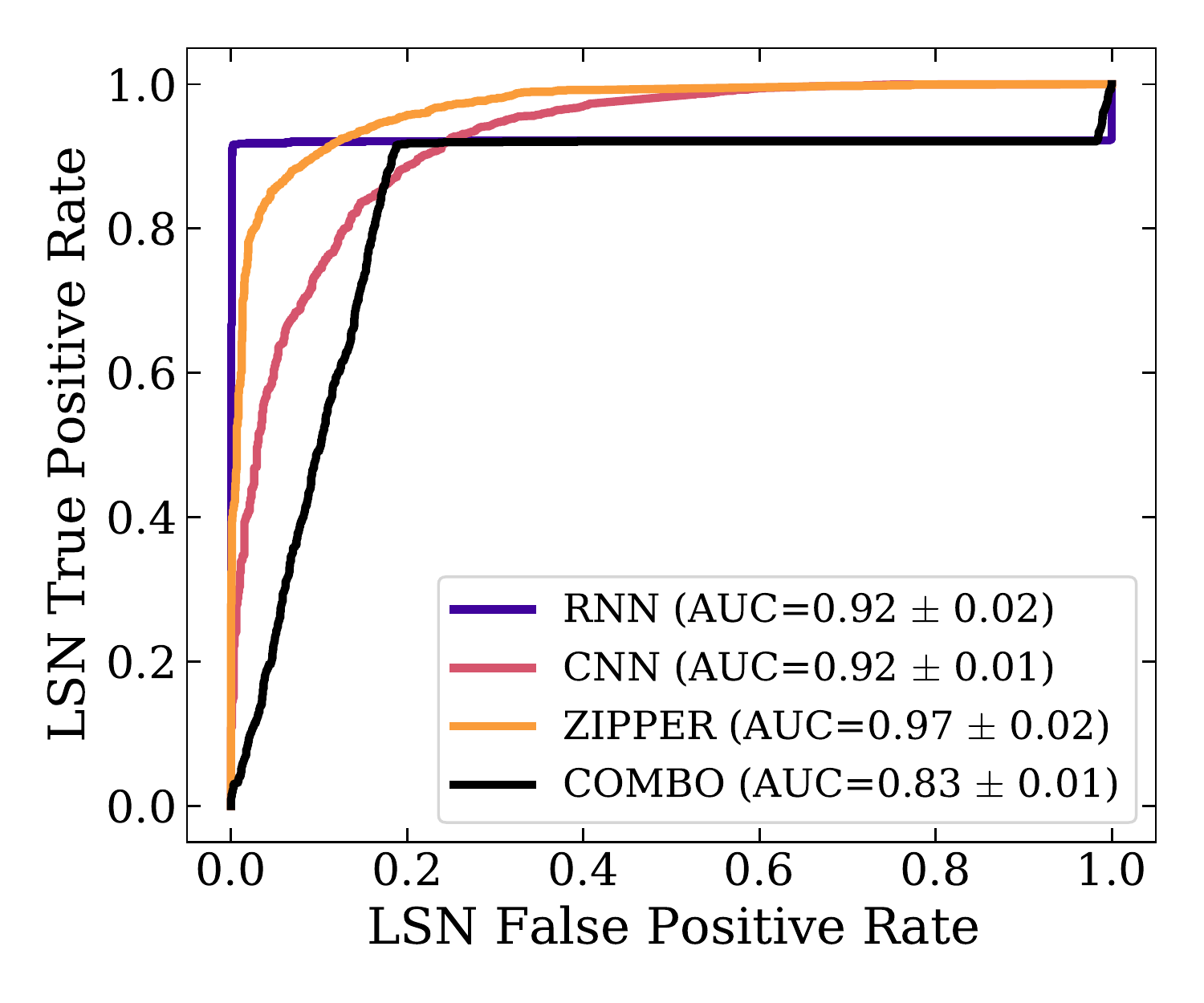}
    \caption{Receiver Operating Characteristic (ROC) curves for the LSST-wide datasets using different networks. The ``COMBO'' label is an optimized weighting of the ``CNN'' and ``RNN'' network predictions. The ROC curves are calculated for the two-class problem of LSNIa and LSNCC versus No Lens and Lens. An Area Under Curve (AUC) of 1.0 indicates perfect performance, while an AUC of 0.5 indicates random guessing.}
    \label{fig:roc_compare}
\end{figure}

Finally, we compare the \ZipperNet\ architecture to a standalone ``RNN,'' a standalone ``CNN,'' and a combination of those two standalone networks (``COMBO'') in the context of the two-class problem and the LSST-wide dataset.
The standalone networks are identical in structure to the corresponding constituents of \ZipperNet\ and trained under the same conditions.
The combination classifier does not connect the feature representations of the standalone networks internally: it merely uses the outputs from each of the standalone classifiers, requiring them to individually identify an object as a LSN.
Therefore, the COMBO classifier reflects a simplistic approach to LSN identification in astronomical surveys --- first a CNN is used to identify a SL system followed by a transient detection algorithm used to search for LSNe.
We expect that the standalone CNN will identify systems with lensing, while the standalone RNN will identify systems with SNe: by requiring both a lensing classification and a SN classification, we establish a baseline for the performance of deep learning architectures that are not connected internally in ways similar to \ZipperNet's design.
ROC curves for the different classifiers are shown in Figure \ref{fig:roc_compare}.
The CNN performs well and mostly identifies all systems with lensing or point sources in galaxies, but this performance can still result in high numbers of false positives (due to the rarity of LSNe) in practical applications.
The RNN is effectively a SN identifier in this context:  the vast majority of the false positives come from classifying unlensed SNe (labels 16 and 17) as LSNe. 
While the ROC AUC is high for the RNN, it alone could not be used as a LSN finder due to the much larger volumetric rate of SNe compared to LSNe.
As shown by this test, \ZipperNet\ outperforms both constituent networks (CNN and RNN) and more importantly the simplistic combination of its constituent networks' outputs (COMBO), indicating that connecting the feature representations of the RNN and CNN internally yields better overall performance and is a worthwhile deep learning strategy for the problem of LSNe detection.
%By itself, a CNN can perform well finding SL systems; by itself a RNN can perform well finding SNe and classifying by type; however, when combined in a united deep learning framework, the resulting architecture is more powerful than either a RNN or a CNN independently.
The \ZipperNet\ architecture shows the highest performance. 

The motivating challenge for \ZipperNet\ is to facilitate the identification of LSNe during the Rubin Observatory's LSST, which is expected to have a very high data stream rate compared to previous large-scale surveys.
The high data stream rate is not a challenge for ZipperNet because classifications can be parallelized and preprocessing is economical (because we  directly extract band-wise lightcurves from  images without the need for deblending or expensive photometric analysis).
The role of a tool like ZipperNet in this setting would be to process the data stream and report a list of candidates ordered by a probability of being a LSN.
At present, the binary classification produced by ZipperNet with confidence quantified by the measured true and false positive rates is our focus.
However, a small amount of additional calibration could straightforwardly map the ZipperNet outputs to physical probabilities to facilitate the generation of candidate lists.

The remaining test of \ZipperNet\ for Rubin Observatory main survey operations is detection at various epochs into the SN light curves for community alerts: How  early into the light curve can \ZipperNet\  find a LSN?
% To quantify the performance in the rapid-response regime,
We present in Figure \ref{fig:discovery} the LSN true-positive rates as functions of the number of lightcurve epochs after the first detection.
In this context, an $r$-band magnitude brighter than 24.4~mag serves as the first detection, which is motivated by the LSST science requirements \citep{lsstsrd}.
We find that even when only one or two epochs are present in the lightcurve after the first detection, the true-positive rate is $> 75\%$, and it improves as more epochs are added.
After five post-detection observing epochs in the light curve, the LSN identification has a true-positive rate $> 95\%$.
Furthermore, even LSNe fainter than the $5\sigma$ limiting magnitude are identifiable with a true-positive rate of $\sim80\%$, indicating high performance where other detection methods would lose sensitivity: both proposed methods discussed in Section \ref{sec:introduction} rely on SNe being brighter than the detection threshold and realized as individual objects.
For these faint LSNe, \ZipperNet\ likely finds success by identifying systems with image-based strong lensing features and then noticing small changes in brightness.
This result supports the expectation that \ZipperNet\ will perform well as a real-time LSN identifier in Rubin Observatory main survey data.

In summary, \ZipperNet\ introduces a new deep learning architecture for lensed transient detection where spatial and temporal features are treated on the same footing in a single framework.
This balanced framework produces a ROC curve AUC of 0.97 when identifying LSNe and $79\%$ accuracy at outright identification of LSNe-Ia in LSST wide-field data even in the early phases of the lightcurve.
Therefore, we expect ZipperNet to play a large role in the rapid identification of LSNe for spectroscopic characterization and time-delay cosmography during the main survey operations of the Rubin Observatory.

%% file: conclusion.tex
\begin{figure}
    \centering
    \includegraphics[trim={0.5cm 0.5cm 0cm 0cm},clip,width=0.47\textwidth]{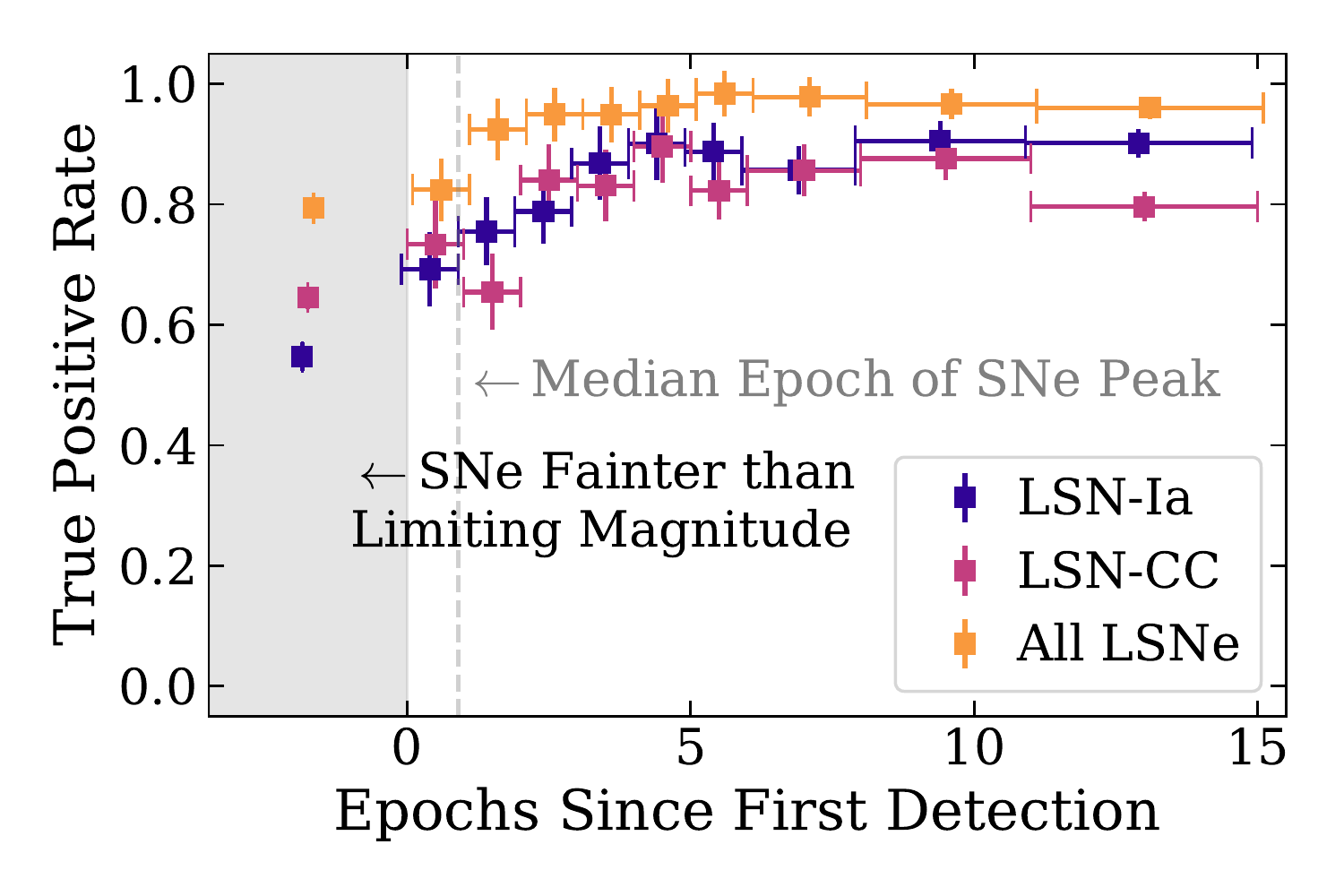}
    \caption{The true-positive rates for LSNe-Ia and LSNe-CC as functions of lightcurve phase for the LSST-wide dataset. 
    The $5\sigma$ limiting magnitude used for the first detection is 24.4 mag in the $r$ band based on the single-epoch survey requirements \citep{lsstsrd}. 
    The ``All LSNe'' label refers to the two-class problem of general LSNe detection.}
    \label{fig:discovery}
\end{figure}

Detecting LSNe soon after explosion will be an important goal for the Vera C. Rubin Observatory and other high-cadence optical surveys.
In this work, we introduced \ZipperNet , a deep learning tool for LSNe identification.
\ZipperNet\ combines a convolutional neural network with a recurrent neural network to simultaneously process spatial and temporal data.
We utilized \texttt{deeplenstronomy} to simulate four distinct optical survey datasets for testing.
\ZipperNet\ performed well when the cadence and seeing were LSST-like or better.
Specifically, \ZipperNet\ was able to identify LSNe in LSST-like data with a ROC AUC of 0.97 and distinguish LSNe-Ia from LSNe-CC in LSST-like data with $79\%$ accuracy.
With \ZipperNet, high-cadence optical surveys can accurately identify both LSNe-CC and LSNe-Ia early into their light curves, and furthermore distinguish between the two transient classes.
Thus, we expect ZipperNet to be a powerful tool in identifying lensing systems for time-delay cosmography measurements during the Rubin Observatory main survey operations.

%% file: acknowledgements.tex
According to the CRediT system we acknowledge the contributions from authors in detail. 
R.~Morgan designed the analysis and deep learning methodologies, prepared simulations, trained networks, and prepared this publication.
B.~Nord and K.~Bechtol offered feedback and guidance on the analysis and publication throughout the project.
S.~J.~Gonzalez prepared merged SDSS / DES catalogs essential for realistic dataset simulation.
L.~Buckley-Geer, A.~M{\"o}ller, J.~W.~Park, and A.~G.~Kim served as internal reviewers within DES.
S.~Birrer provided useful comments on the draft and analysis.
All other authors contributed to DES infrastructure upon which this project was based.

R.~Morgan thanks the Universities Research Association Fermilab Visiting Scholars Program for funding his work on this project. 
R.~Morgan also thanks the LSSTC Data Science Fellowship Program, which is funded by LSSTC, NSF Cybertraining Grant \#1829740, the Brinson Foundation, and the Moore Foundation; his participation in the program has benefited this work. 

We acknowledge the Deep Skies Lab as a community of multi-domain experts and collaborators who’ve facilitated an environment of open discussion, idea-generation, and collaboration. This community was important for the development of this project.

This material is based upon work supported by the National Science Foundation Graduate Research Fellowship Program under Grant No. 1744555. Any opinions, findings, and conclusions or recommendations expressed in this material are those of the author(s) and do not necessarily reflect the views of the National Science Foundation.

 Funding for the DES Projects has been provided by the U.S. Department of Energy, the U.S. National Science Foundation, the Ministry of Science and Education of Spain, 
 the Science and Technology Facilities Council of the United Kingdom, the Higher Education Funding Council for England, the National Center for Supercomputing 
 Applications at the University of Illinois at Urbana-Champaign, the Kavli Institute of Cosmological Physics at the University of Chicago, 
 the Center for Cosmology and Astro-Particle Physics at the Ohio State University,
 the Mitchell Institute for Fundamental Physics and Astronomy at Texas A\&M University, Financiadora de Estudos e Projetos, 
 Funda{\c c}{\~a}o Carlos Chagas Filho de Amparo {\`a} Pesquisa do Estado do Rio de Janeiro, Conselho Nacional de Desenvolvimento Cient{\'i}fico e Tecnol{\'o}gico and 
 the Minist{\'e}rio da Ci{\^e}ncia, Tecnologia e Inova{\c c}{\~a}o, the Deutsche Forschungsgemeinschaft and the Collaborating Institutions in the Dark Energy Survey. 

The Collaborating Institutions are Argonne National Laboratory, the University of California at Santa Cruz, the University of Cambridge, Centro de Investigaciones Energ{\'e}ticas, 
Medioambientales y Tecnol{\'o}gicas-Madrid, the University of Chicago, University College London, the DES-Brazil Consortium, the University of Edinburgh, 
the Eidgen{\"o}ssische Technische Hochschule (ETH) Z{\"u}rich, 
Fermi National Accelerator Laboratory, the University of Illinois at Urbana-Champaign, the Institut de Ci{\`e}ncies de l'Espai (IEEC/CSIC), 
the Institut de F{\'i}sica d'Altes Energies, Lawrence Berkeley National Laboratory, the Ludwig-Maximilians Universit{\"a}t M{\"u}nchen and the associated Excellence Cluster Universe, 
the University of Michigan, NSF's NOIRLab, the University of Nottingham, The Ohio State University, the University of Pennsylvania, the University of Portsmouth, 
SLAC National Accelerator Laboratory, Stanford University, the University of Sussex, Texas A\&M University, and the OzDES Membership Consortium.

Based in part on observations at Cerro Tololo Inter-American Observatory at NSF’s NOIRLab (NOIRLab Prop. ID 2012B-0001; PI: J. Frieman), which is managed by the Association of Universities for Research in Astronomy (AURA) under a cooperative agreement with the National Science Foundation.

 The DES data management system is supported by the National Science Foundation under Grant Numbers AST-1138766 and AST-1536171.
 The DES participants from Spanish institutions are partially supported by MICINN under grants ESP2017-89838, PGC2018-094773, PGC2018-102021, SEV-2016-0588, SEV-2016-0597, and MDM-2015-0509, some of which include ERDF funds from the European Union. IFAE is partially funded by the CERCA program of the Generalitat de Catalunya.
 Research leading to these results has received funding from the European Research
 Council under the European Union's Seventh Framework Program (FP7/2007-2013) including ERC grant agreements 240672, 291329, and 306478.
 We  acknowledge support from the Brazilian Instituto Nacional de Ci\^encia
 e Tecnologia (INCT) e-Universe (CNPq grant 465376/2014-2).

This paper has gone through internal review by the DES collaboration.

This manuscript has been authored by Fermi Research Alliance, LLC under Contract No. DE-AC02-07CH11359 with the U.S. Department of Energy, Office of Science, Office of High Energy Physics.